\def\la{\mathrel{\hbox{\rlap{\hbox{\lower4pt\hbox{$\sim$}}}\hbox{$<$}}}}
\def\ga{\mathrel{\hbox{\rlap{\hbox{\lower4pt\hbox{$\sim$}}}\hbox{$>$}}}}

\documentclass{iau} 
\usepackage{graphicx}

\title[X-ray - IR relation of AGNs in NEP] %% give here short title %%
{X-ray - Infrared relation of AGNs and search for highly obscured accretion in the {\sl AKARI} NEP Field}

\author[T. Miyaji \& {\sl AKARI} NEP Survey Team]   %% give here short author list %%
{Takamitsu Miyaji$^1$
 \and AKARI NEP Survey Team}

\affiliation{$^1$Instituto de Astronom\'ia, Universidad Nacional Aut\'onoma de M\'exico,
Km. 103 Carret. Tijuana-Ensendada, Ensenada, 22860 Mexico 
\\ email: {\tt miyaji@astro.unam.mx} \\[\affilskip]}

\pubyear{2019}
\volume{341}  %% insert here IAU Symposium No.
\jname{PanModel2018 : Challenges in Panchromatic Galaxy Modelling with Next Generation Facilities}
\editors{M. Boquien, E. Lusso, C. Gruppioni, and P. Tissera eds.}
\begin{document}

\maketitle

\begin{abstract}

The infrared Astronomical Satellite {\it AKARI} conducted deep ($\sim$ 0.4 deg$^2$) 
and wide ($\sim 5.4$ deg$^2$) surveys around the North Ecliptic Pole (NEP) with its
InfraRed Camera (IRC) with nine filters continuously covering the 2-25 $\mu$m range.
These photometric bands include three filters that fill the ``{\it Spitzer} gap'' between 
the wavelength coverages of IRAC and MIPS. This unique feature has enabled us 
to make sensitive mid-infrared detection of AGN candidates at z$\sim$ 1-2,
based on the Spectral Energy Distribution (SED) fitting including hot dust emission in the 
AGN torus. This enables us to compare X-rays and the AGN torus component of the infrared 
emission to help us identify highly absorbed AGNs, including Compton-thick ones.  
We report our results of the {\it Chandra} observation of the {\sl AKARI} NEP Deep Field 
and discuss the prospects for upcoming {\it Spectrum-RG} (eROSITA+ART-XC) on the 
{\sl AKARI} Wide field. 
\end{abstract}

\firstsection % if your document starts with a section,
\section{Introduction}

 There are three major bumps in the electromagnetic continuum in a typical 
AGN: 1) optically-thick thermal emission from the accretion disk in the ultraviolet, 2) the primary
X-ray continuum  by the inverse-Compton scattering of the disk ultraviolet photons 
by hot coronae, and 3) the infrared emission from hot dust in the circum-nuclear torus. Additionally,
reprocessed primary X-ray continuum by reflections at surrounding material produces a
characteristic ``Comption hump'' in hard X-rays. Observational determination of how these three 
components are inter-related is a key to panchromatic modelings of the AGN component in galaxies. 

%Down to a rather low X-ray luminosity ($L_{\rm X}\ga 10^{42}{[\rm erg\,s^{-1}]}$), 
%most of the X-ray emission of the galaxy is likely  to originate from the AGN, with very little 
%contamination from stellar/star-formation components. 
In multiwavelength extragalactic surveys, X-rays are considered  the most efficient marker 
of the AGN activity. However it is subject to absorption by intervening gas by photo-electric absorption, 
which occurs mainly at photon energies $E<10$ keV. If the line of sight column density becomes 
$N_{\rm H}\ga 10^{24}{\rm cm^{-2}}$, the effects Compton scattering further attenuates 
X-ray emission even at $E\ga 10$ keV. 

While the infrared emission from hot dust heated by AGNs is less subject to absorption by
intervening material,  it has to be separated from stars and warm dust from star-formation regions
by an SED decomposition to measure the IR luminosity of the AGN component alone.
 
The continuous 9-band 2-25 $\mu$m photometric data available 
from {\sl AKARI} NEP Survey are ideal for such SED decomposition of galaxies. 
The availability of three photometric bands in the 11-20 $\mu$m range, which is
the wavelength gap between IRAC and MIPS instruments of {\it Spitzer},  
makes the {\sl AKARI} IRC dataset a very strong tool for separating 
AGN hot dust emission from stellar+star-formation components at $1\la z \le 2$ 
(the ``Cosmic Noon'').   

In this proceedings article, we summarize the results and prospects of our X-ray followup 
studies of the {\sl AKARI} NEP field with emphasis on X-ray data and the connection between
X-rays and infrared. These include our {\it Chandra} followup studies of the {\sl AKARI} NEP 
Deep Field and prospects of upcoming {\it Spectrum-RG} (eROSITA+ART-XC)  observations.  
    
\section{The {\it AKARI} NEP Survey}

{\sl AKARI} was a Japanese IR astronomical satellite (\cite[Matsuhara et al. 2005]{matsuhara05};
see also Goto et al., this symposium), which was operated from 2006 to 2011. 
%\sout{In addition to the 
%well-known mid- to far-IR all-sky surveys (Ishihara et al.~2010), 
%AKARI performed spectroscopic surveys and deep imaging surveys in 13
%bands ranging from 2--160 $\mu$m, as well as pointed observations.}
One of the unique features of the observatory was the availability
of continuous wavelength coverage over 2--25 $\mu$m in 9 filters with
its IR Camera (IRC), including three filters in the 11--19 $\mu$m range. 
These  filters fill the 9--20 $\mu$m gap between the Spitzer IRAC and MIPS instruments and allows 
an efficient selection of AGN at $1 \la z \la 2$
in the MIR. This redshift range (the so-called Cosmic Noon) is particularly important to study 
since that is where both the X-ray detected number and luminosity densities peak. 
% This epoch may well dominate the SMBH growth. 
%It is therefore very important to search for and quantify the obscured accretion 
%in this redshift range, where the use of {\sl Spitzer} is limited. 

\begin{figure}
\begin{center}
 \includegraphics[width=0.55\textwidth]{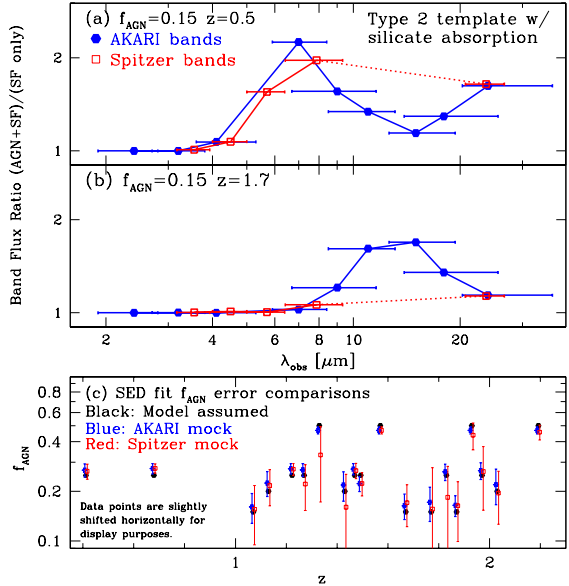}
% \vspace*{-0.5 cm}
 \caption{{\sl (a) \& (b):} Excess MIR emission ([AGN+Starburst]/Starburst),
due to the presence of a type 2 AGN (with the 9.7 $\mu$m silicate absorption feature) 
observed through the {\sl AKARI} IRC (blue/filled hexagon) and {\sl Spitzer} IRAC+MIPS (red/open square) 
filter sets, for (a) $z=0.5$ and (b) $z=1.7$ objects. An AGN fraction 
$f_{\rm AGN}=0.15$ of the total IR luminosity is assumed.
{\sl (c):} Comparisons of expected $f_{\rm AGN}$ errors between SED fits to 
a number of simulated SEDs (black/filled symbol without error bars) through {\sl AKARI}+{\sl Herschel}
(blue/filled symbol with error bars) and 
{\sl Spitzer}+{\sl Herschel} (red/open square with error bars). In calculating the errors, imaging depths of 
ANEPD and COSMOS are assumed for the former and the latter respectively. The calculations
are made by the Cigale software using the model in \cite[Ciesla et al. (2015)]{ciesla2015}. \label{akari_vs_spitzer}}
\end{center}
\end{figure}

 Figure \ref{akari_vs_spitzer} demonstrates the advantage of the continuous wavelength 
coverage of {\sl AKARI} 9-band photometry (including 10--19 $\mu$m) over the {\it Spizer} 
IRAC+MIPS photomtetric bands in detecting the AGN torus component. (See caption.)

As a legacy program of {\sl AKARI}, a major portion of all pointed observations
during its liquid helium phase was devoted to surveys around North Ecliptic Pole (NEP),
namely the AKARI NEP Deep ($\sim 0.4$ deg$^2$; hereafter ANEPD) and Wide
($\sim 5.4$ deg$^2$; ANEPW) field surveys with IRC.  At ANEPD, the {\sl AKARI} data have 5$\sigma$ sensitivity 
limits of, e.g.,  $\sim$9 and $87$ $\mu$Jy at 3.2 and 15 $\mu$m respectively 
(\cite[Murata et~al. 2013]{murata13}), while the limiting fluxes are approximately 
twice as large in NEPW (\cite[Kim et~al. 2012]{kim12}).

\section{Chandra Observations of the {\it AKARI} NEP Deep Field}

We were granted a 250 ks dense tiling of a 3 $\times$ 4 \textit{Chandra} ACIS-I mosaic survey
in the ANEPD field (\cite[Krumpe et al. 2015]{krumpe15}; hereafter K15).
Including additional  $\sim 50$ ks of data from archive, we cover the field  with
exposure times ranging from $\sim$ 40 -- 80 ks. The source catalog and 
the first analysis of 457 X-ray sources is presented in K15. 

\begin{figure}
\begin{center}
 \includegraphics[width=0.45\textwidth]{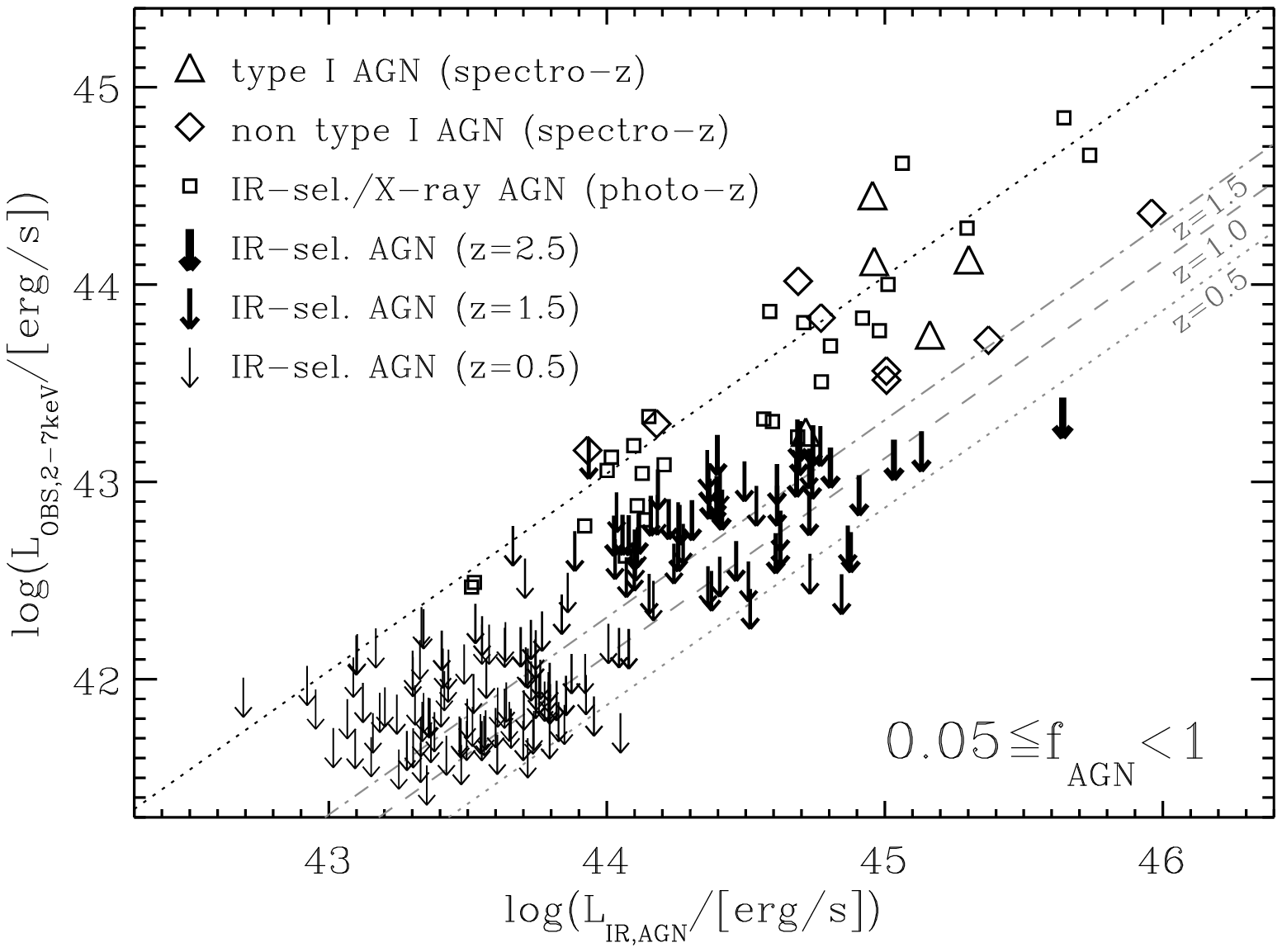} 
 \includegraphics[width=0.45\textwidth]{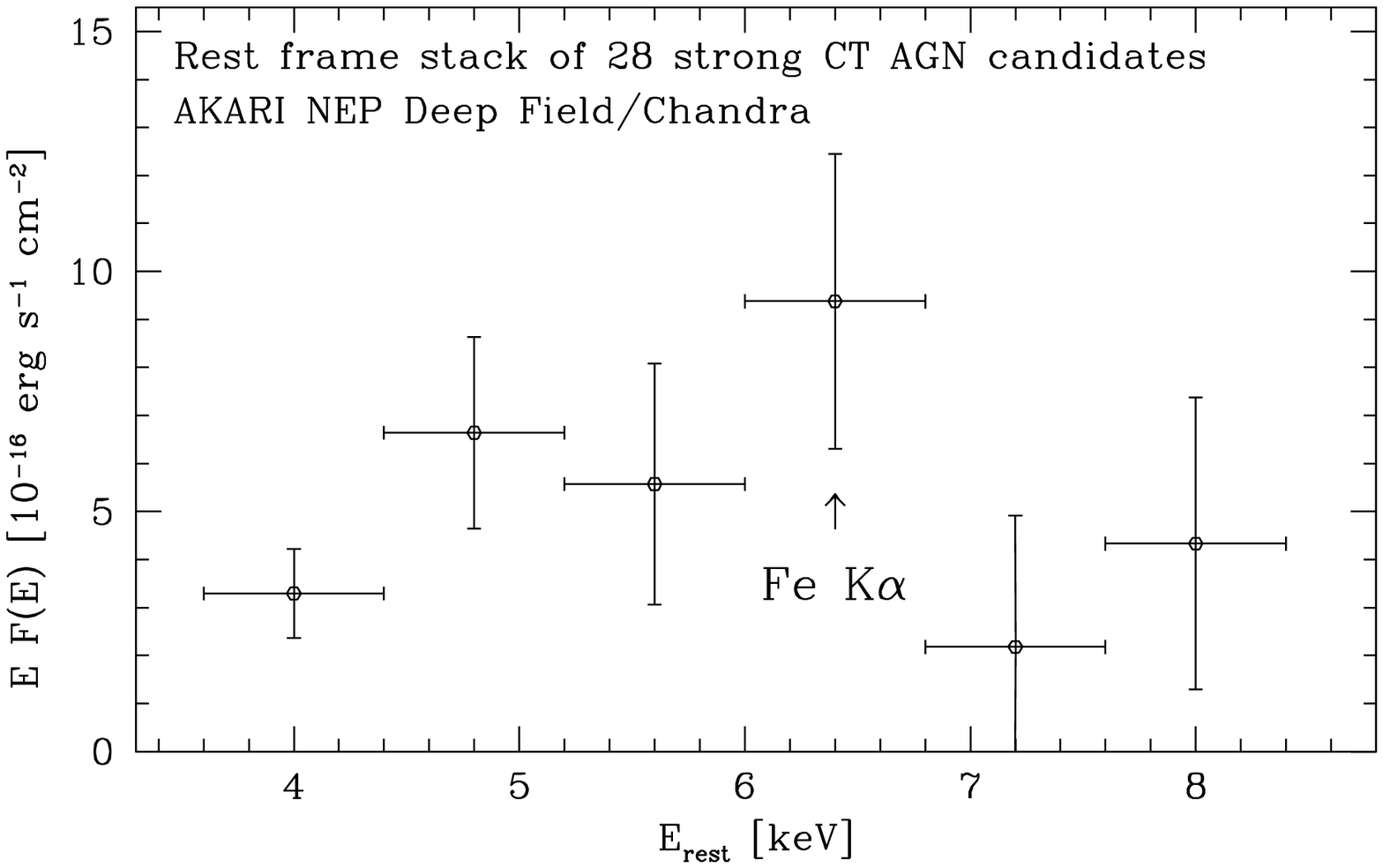} 
% \vspace*{-1.0 cm}
 \caption{{\it Left:} Infrared AGN rest-frame luminosity from SED fits vs. hard 
               (2--7 keV) observed X-ray luminosity. The symbols represent: triangles -- spectroscopically-confirmed 
               type I AGNs, diamonds -- objects with optical spectra that do not show broad emission lines,
               small squares -- IR-selected AGN that are also detected in the 2--7 keV band with 
               photometric redshifts only, downward arrows -- IR-selected AGN undetected 2--7 keV band,
               where the vertical position corresponds to the 2-7 keV flux upper limit. The thickness 
               of the arrows represents the redshift of the source. 
               The dotted line shows the correlation between the AGN IR luminosity 
               and the X-ray luminosity ($\langle L_{\rm X}/L_{\rm IR}\rangle = 0.11$) for unabsorbed AGNs. 
               The gray lines show for different redshifts the expected attenuation of the 
               2--7 keV X-ray luminosity caused by an intrinsic absorption of 
               $N_{\rm H} = 10^{24}$ cm$^{-2}$. Adopted from K15 (Fig. 18).
               {\it Right:} The rest-frame stacked spectrum of 
               the 28 strong CT AGN candidates calculated using the CSTACK utility.}
   \label{fig:chandra}
\end{center}
\end{figure}

 In K15, (see Fig. \ref{fig:chandra} {\it left}), we compare the observed X-ray (2-7 keV) luminosity $L_{\rm OBS,2-7 keV}$, i.e. 
 luminosity calculated without K-correction or absorption correction) and the AGN component of the IR luminosity
 $L_{\rm IR,AGN}$ obtained from SED fitting using an updated version of the
 \cite[Hanami et al. (2012)]{hanami12}'s procedure. By this comparison,
we selected 28 strong Compton-thick AGN (CTAGN) candidates, i.e., AGNs with a line of sight column 
density of  $N_{\rm H}>10^{24}{\rm [cm^{-2}]}$. 

 The CTAGNs are characterized by a large equivalent width of the Fe K$\alpha$ line ($EW\ga 1$ keV). 
Using the CSTACK utility\footnote{http://lambic.astrosen.unam.mx/cstack}, we have 
made a rest-frame stacking of the X-ray spectra of the 28 strong CT AGN candidates (Fig. \ref{fig:chandra} {\it right}), 
obtaining an equivalent width of the Fe K$\alpha$ of $EW=1.0\pm 0.6$ keV. This
suggests that a majority of these candidates are CTAGNs with possible contamination from non-CT AGNs. 

\section{Prospects for eROSITA/ART-XC observations}

The spacecraft {\it Spectrum-RG}, which is planned to be launched in the near future
(Merloni, this symposium), carries two instruments, eROSITA (sensitive in 0.2-10 keV) and ART-XC (5-30 keV).
The plan of the mission is to perform 8 all-sky surveys over four years and the every great circle scan will go through the Ecliptic pole regions.
{\it Spectrum-RG} will provide with us the best-matched dataset for our studies of highly obscured AGNs
in the $\sim 5.4$ deg$^2$ ANEPW field, which is centered at the exact NEP. The combination of the ANEPW survey with the
eROSITA dataset, we expect to probe a high luminosity - highly absorbed AGN population.  

 Using the current determination of the X-ray luminosity function of AGNs at $0.4<z<2$ (\cite[Miyaji et al. 2015]{miyaji15}),
$N_{\rm H}$ function (\cite[Ueda et al. 2014]{ueda14}), and the expected flux limit of eROSITA in the 2-10 keV band 
(\cite[Merloni et al. 2012]{merloni12}), we expect to identify $\sim 270$ CTAGN/semi-CTAGN 
 candidates using the method described above. The ART-XC instrument is sensitive in 5-30 keV, we expect to improve the distinction 
between non-CT AGNs and CT-AGNs by adding the ART-XC dataset.

\section{Concluding Remarks}

We are in the process of improving the investigation. The improvements include: 
\begin{itemize}
\item Spectroscopic followup of ANEPD X-ray sources, IR-selected AGNs and  and CTAGN candidates, using large telescopes including Gran Telescopio Canarias, Keck, and Large Binocular Telescope. 
%This will identify more broad-line (type I AGNs), type II AGNs from the emission-line diagnostics  and will provide accurate redshifts for more accurate  SED fittings and rest-frame X-ray stackings. 
\item We will use updated models for the AGN torus. These include the Clumpy Torus model (\cite[Nenkova et al. 2015]{nenkova15}) on the 
infrared side. An X-ray torus clumpy model that has the consistent geometrical configurations/parameter definitions as the \cite[Nenkova et al. (2015)]{nenkova15} model is being developed by a group at Kyoto University (Tanimoto et al.,Ogawa et al. this symposium). 
We will use these models for  a consistent IR-X-ray modeling in selecting CT-AGNs. 
\end{itemize}

TM is supported by UNAM-DGAPA PAPIIT (IN104216, IN111319) and CONACyT (252531). He thanks the conference organizers/NAOJ for a travel support.

\begin{discussion}
\discuss{Triani}{Is there any effect of inclination angle in the AGN SED in IR?}
\discuss{Miyaji}{Yes, there is. Especially, there is silicate absorption in the case of an edge-on view of the torus.
We are investigating in detail the IR SED of tori including the angle dependence of the Nenkova ``Clumpy Torus'' model. Also 
on the X-ray side, we will revise the model using the new ``X-clumpy'' torus model, which has matching  geometries and parameter definitions
of the Nenkova model. This is under development by a group at Kyoto University and some results of its preliminary version of
are presented in this conference (posters P36 by Tanimoto and P22 by Ogawa).} 
\end{discussion}

\end{document}